\documentclass[preprint2,times,tighten]{aastex6}
\usepackage{amsmath,amstext}
\usepackage[figure,figure*]{hypcap}
\usepackage{newtxmath} 
\usepackage[utf8]{inputenc}
\usepackage[T1]{fontenc}
\shorttitle{Circular polarization disturbances in a sunspot}
\shortauthors{M. Stangalini et al.}
\newcommand{\angs}{\text{\normalfont\AA}}

\begin{document}

\title{Propagating Spectropolarimetric Disturbances in a Large Sunspot}
\author{M. Stangalini\altaffilmark{1}, S. Jafarzadeh\altaffilmark{2,3}, I. Ermolli\altaffilmark{1}, R. Erdélyi\altaffilmark{4,5}, D. B. Jess\altaffilmark{6,7}, P. H. Keys\altaffilmark{6}, F. Giorgi\altaffilmark{1}, M. Murabito\altaffilmark{1}, F. Berrilli\altaffilmark{8}, D. Del Moro\altaffilmark{8}}

\email{marco.stangalini@inaf.it}
\altaffiltext{1}{INAF-OAR National Institute for Astrophysics, Via Frascati 33, 00078 Monte Porzio Catone (RM), Italy}
\altaffiltext{2}{Rosseland Centre for Solar Physics, University of Oslo, P.O. Box 1029 Blindern, NO-0315 Oslo, Norway}
\altaffiltext{3}{Institute of Theoretical Astrophysics, University of Oslo, P.O. Box 1029 Blindern, NO-0315 Oslo, Norway}
\altaffiltext{4}{Solar Physics \& Space Plasma Research Centre (SP2RC), School of Mathematics and Statistics, University of Sheffield, Sheffield S3 7RH, UK}
\altaffiltext{5}{Department of Astronomy, E\"otv\"os L. University, P\'azm\'any P. s\'et\'any 1/A, Budapest H-1117, Hungary}
\altaffiltext{6}{Astrophysics Research Centre, School of Mathematics and Physics, Queen’s University Belfast, Belfast BT7 1NN, UK}
\altaffiltext{7}{Department of Physics and Astronomy, California State University Northridge, Northridge, CA 91330, USA}
\altaffiltext{8}{Department of Physics, Università di Roma Tor Vergata, Via della Ricerca Scientifica 1, 00133, Rome, Italy}

\begin{abstract}
We present results derived from the analysis of spectropolarimetric measurements of active region AR12546, which represents one of the largest sunspots to have emerged onto the solar surface over the last $20$ years. The region was observed with full-Stokes scans of the Fe I 617.3 nm and Ca II 854.2 nm lines with the Interferometric BIdimensional Spectrometer (IBIS) instrument at the Dunn Solar Telescope over an uncommon, extremely long time interval exceeding three hours. Clear circular polarization (CP) oscillations localized at the umbra-penumbra boundary of the observed region were detected. Furthermore, the multi-height data allowed us to detect the downward propagation of both CP and intensity disturbances at $2.5-3$~mHz, which was identified by a phase delay between these two quantities. These results are interpreted as a propagating magneto-hydrodynamic surface mode in the observed sunspot.
 
\end{abstract}

\section{Introduction}
Driven by the forcing action of steady and impulsive photospheric plasma motion, a large variety of magneto-hydrodynamic (MHD) modes (e.g. kink, torsional Alfvén, sausage) can be excited in sunspots and small scale magnetic field concentrations \citep[e.g.,][to mention a few]{Edwin1983, Roberts1983, Khomenko2008, 2010ApJ...719..357F, 2011ApJ...727...17F, 2015MNRAS.449.1679M, 2015ApJ...799....6M, 2016PhDT........15L, Grant:2018vfa, 2018ApJ...857...28K}.
In addition to this, the magnetic field concentrations can interact with the surrounding {\it p-} and {\it f-}mode oscillations, and these can be eventually absorbed and converted into magneto-acoustic waves that can propagate along the field guide to the upper layers of the solar atmosphere \citep[see for example][]{1992ApJ...391L.113B, 1994ApJ...437..505C, 2005SoPh..227....1C, 2007A&A...471..961M, 2012ApJ...755...18V, 2014ApJ...791...61F, 2015ApJ...806..132G, 2017ApJ...842...59J, 2018AdSpR..61..759P}. There is a general consensus that MHD waves can play a significant role in the energy budget of the solar atmosphere \citep{2009Jess, 2011ApJ...735...65F, 2012NatCo...3E1315M, 2013SSRv..175....1M, 2015SSRv..190..103J, 2017ApJ...847....5K}.

  \begin{figure*}
  \centering
    \includegraphics[trim=0.5cm 0cm 0.cm 0cm, clip, width=17cm]{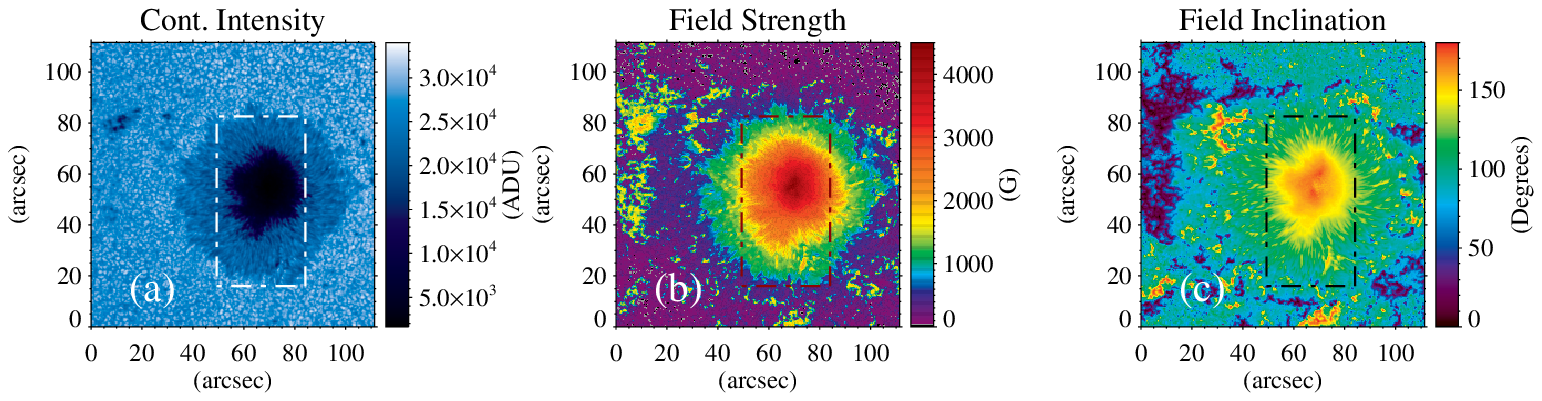}
  \includegraphics[trim=0.5cm 0cm 0.cm 0cm, clip, width=18.cm]{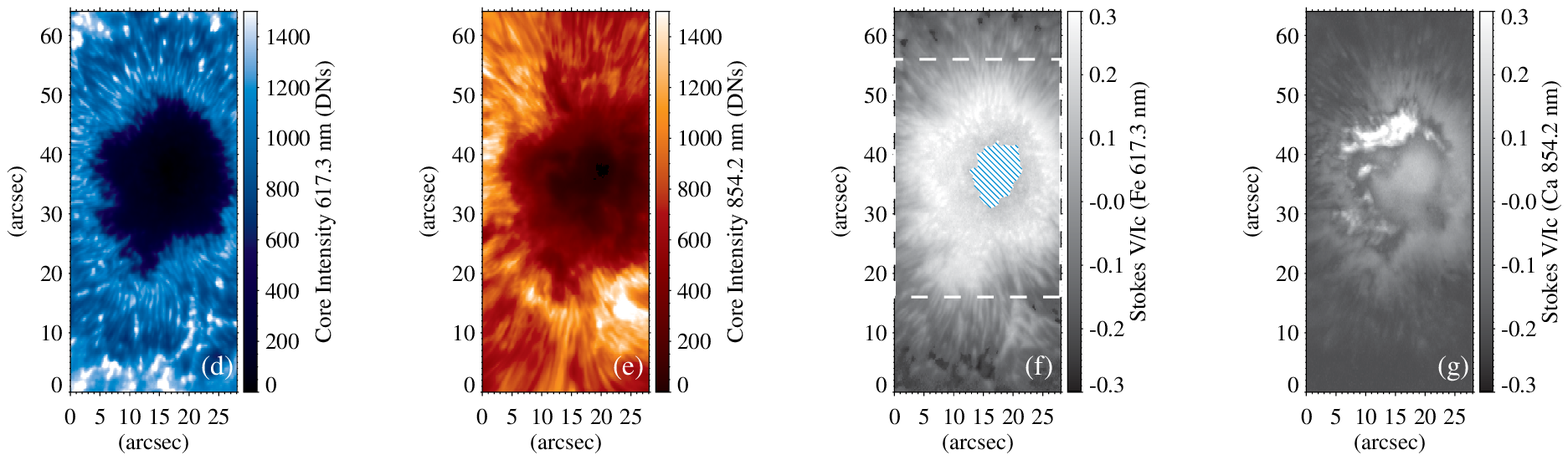}
  \caption{SOT/SP continuum intensity (panel a), magnetic field strength (panel b) and inclination (panel c) from HINODE level 2 data products. The dot-dashed rectangular box represents the IBIS FoV. (Panels d and e): IBIS intensity images in the core of the Fe I 617.3 nm line and of the  Ca II $854.2$ nm  line, respectively. (Panel f): $CP$ map obtained from the IBIS Fe I $617.3$ nm line measurements. The central region of the umbra that is affected by possible saturation effects and low photon flux is indicated as a hatched area. (Panel g): $CP$ map derived from the IBIS Ca II  $854.2$ nm  line data.}
  \label{Fig:context}
  \end{figure*}
  
From an observational point of view, a wealth of wave signatures in different observables were reported so far at different heights in the solar atmosphere and in different kinds of magnetic structures, based on intensity and Doppler velocity measurements \citep{2000SoPh..192..373B, 2006ApJ...640.1153C, 2010A&A...513A..27C, 2011ApJ...729L..18M, 2012A&A...539L...4S, 2015ApJ...806..132G, 2017ApJS..229...10J, 2017ApJ...842...59J}. The propagation of these disturbances depends on the local physical parameters and the geometry of the waveguide \citep{2006ApJ...647L..77M, 2006ApJ...648L.151J, 2007ApJ...671.1005B, Stangalini2011, 2015A&A...579A..73M, 2017SoPh..292...35A, 2018ApJ...853..136K}.\\
In addition to intensity and Doppler velocity oscillations, magnetic field perturbations are also expected in the case of particular MHD modes \citep{Edwin1983, Roberts1983}. Several attempts have been made to detect such oscillations in different magnetic structures in the Sun's atmosphere. Many authors have reported magnetic field oscillations with periods in the range $3-5$ min and amplitudes of the order of 10 G \citep[e.g.,][to mention a few]{1997SoPh..172...69H, 1999ASSL..243..337R, 1999SoPh..187..389B, 2002AN....323..317S, 2000ApJ...534..989B, 2018arXiv180300018H}. However, based on the available observations, the unambiguous attribution of these perturbations to magnetic oscillations has been debated \citep[][]{2002AN....323..317S}.\\
Indeed, by making use of spectropolarimetric inversions in the photosphere, \citet{1998ApJ...497..464L} found an upper limit for the amplitude of the magnetic fluctuations in a large, symmetric sunspot located near disc center of the order of $4$ G, and interpreted them of instrumental origin. Independent studies also found that significant magnetic oscillations are inhomogeneously distributed and concentrated in patches \citep{1998A&A...335L..97R}, or at the boundaries between the umbra-penumbra boundary in sunspots \citep{1999SoPh..187..389B, 2000SoPh..191...97K}. Through phase lag analyses between different physical quantities, it was shown that these magnetic perturbations were not consistent with cross-talk between either the Doppler shift of the line, nor the intensity. Based on that, \cite{2000SoPh..191...97K} argued that the observed oscillations of the Stokes parameters in sunspots could be interpreted as real magnetic field perturbations resulting from the swaying of the field lines in response to the driving action of \textit{p}-modes.  
  
  \begin{figure*}
  \centering
  (a)\includegraphics[trim=0.5cm 0cm 0cm 0.6cm, clip, height=6.1cm]{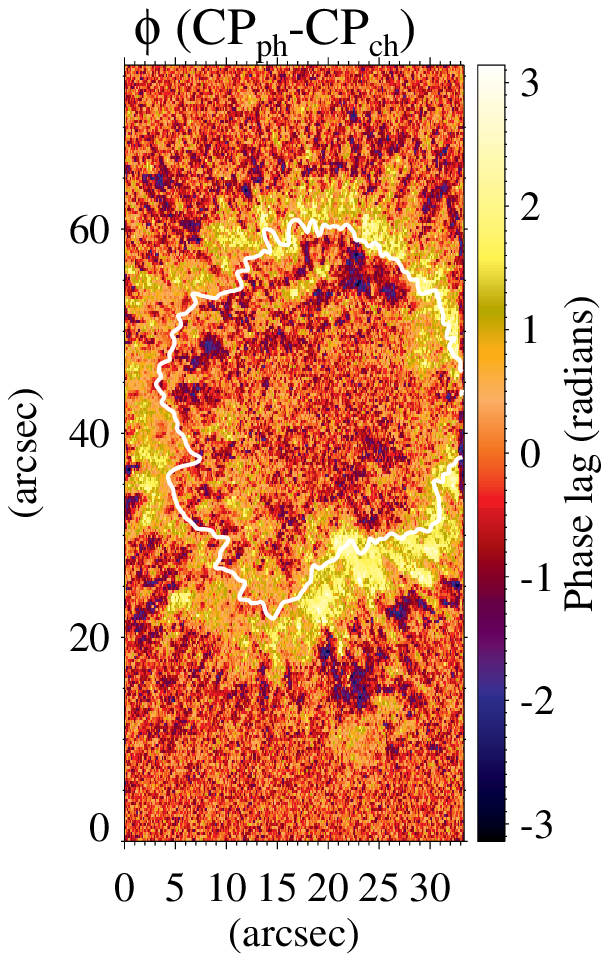}
  (b)\includegraphics[trim=2.1cm 0cm 0cm 0.6cm, clip, height=6.1cm]{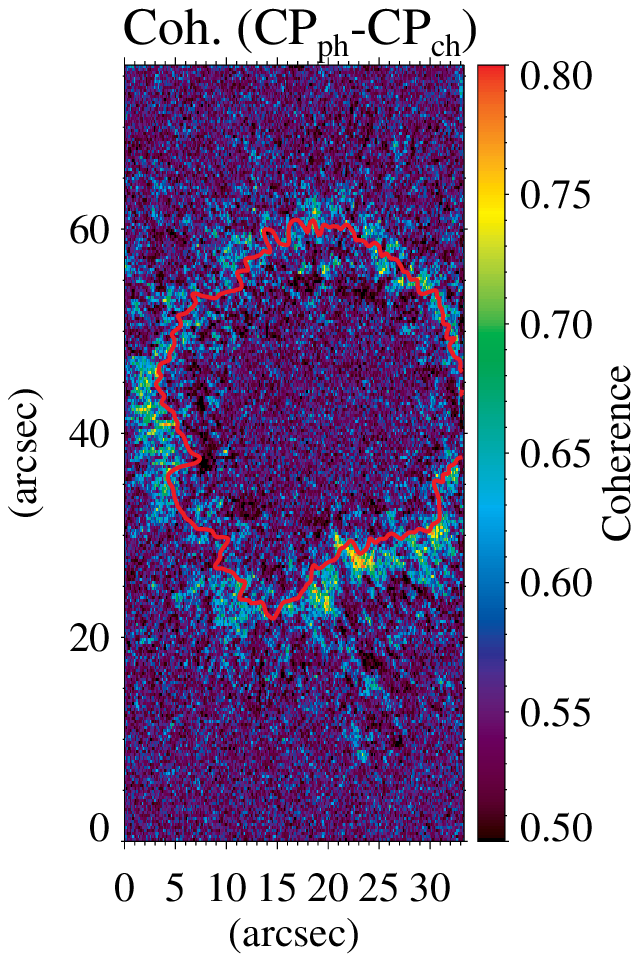}
  (c)\includegraphics[trim=2.1cm 0cm 0cm 0.6cm, clip, height=6.1cm]{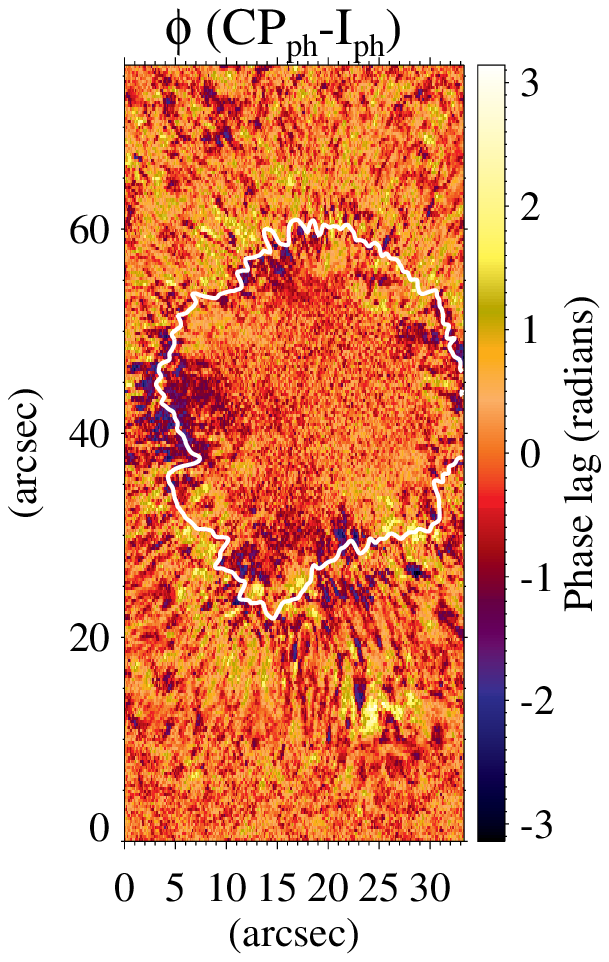} 
  (d)\includegraphics[trim=2.1cm 0cm 0cm 0.6cm, clip, height=6.1cm]{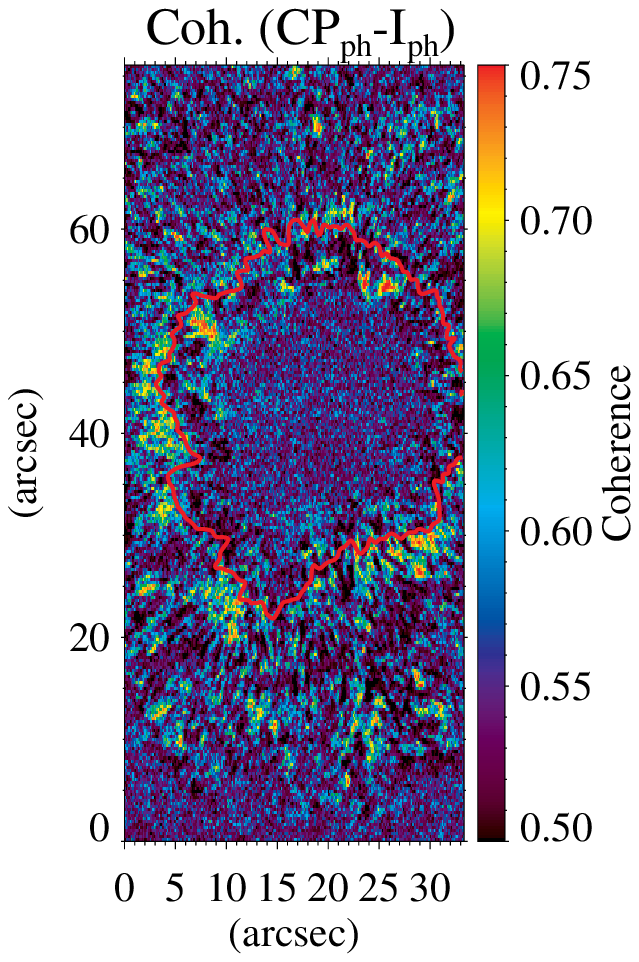} 
  \caption{(a) Phase lag map of $CP$ fluctuations at $3$ mHz (with a bandwidth of 0.7 mHz) between the photosphere and chromosphere. (b) Coherence map at $3$ mHz (with a bandwidth of 0.7 mHz) for the same $CP$ disturbances. (c) Phase lag map at the same frequency band computed between $CP$ fluctuations and core intensity fluctuations in the photosphere. (d) Coherence map corresponding to the phase map of panel (c). The continuous contours indicate the approximate position of the umbra-penumbra boundary as seen in the continuum intensity. The dashed lines highlight the region where the analysis was performed (see text for more details). Please note that the maps are obtained by averaging four spectral bins in Fourier space or, equivalently, $3.0\pm0.7$~mHz. For this reason, the average phase and coherence values might appear lower than they are at each frequency bin.}
  \label{Fig:phasemaps}
  \end{figure*}

\citet{2009ApJ...702.1443F}, using spectropolarimetric data of the solar photosphere acquired by \textit{Hinode} SOT/SP \citep{springerlink:10.1007/s11207-008-9174-z}, reported the presence of magnetic flux oscillations in pores and other magnetic concentrations and interpreted them as the signature of sausage and kink modes. \\
More recently, \citet{2011ApJ...730L..37M} by taking advantage of the unprecedented high spatial resolution achieved by the balloon-borne {\sc Sunrise} mission \citep{2010ApJ...723L.127S}, detected oscillations of the magnetic flux density in small-scale magnetic elements in the quiet Sun. These oscillations, being in antiphase with area oscillations of the magnetic element, were consistent with intrinsic magnetic oscillations of the flux tube. The difficulty of mode identification was addressed by, e.g., \citet{2013A&A...551A.137M}, \citet{2013A&A...555A..75M}, and more recently by \citet{2015A&A...579A..73M}.\\
As mentioned earlier and underlined by \citet{2015LRSP...12....6K}, the detection of magnetic oscillations may suffer from cross-talk with other physical quantities. In addition, in the presence of a vertical gradient of the magnetic field, opacity effects may also play a significant role. In this regard, it is worth underlining that the phase analysis between different quantities can be important to verify the real nature of the detected oscillations in the solar atmosphere \citep[see][]{2015A&A...579A..73M}. This approach was already adopted by some authors \citep{1999SoPh..187..389B, 2000SoPh..191...97K} to discriminate between intrinsic magnetic fluctuations and cross-talks from other physical quantities such as temperature and density fluctuations associated with magneto-acoustic waves \citep{2003A&A...410.1023R}. However, the phase lag analysis was limited to a single photospheric height. Extending the phase lag analysis to spectropolarimetric diagnostics acquired simultaneously at different heights in the solar atmosphere can provide new insight into the nature of these observed magnetic perturbations and, possibly, their propagation. This is the novel aspect of this study and the main scope of this work.
  
\section{Data set and methods}
The data set used in this work was acquired on May 20th 2016 with the Interferometric BIdimensional Spectrometer \citep[IBIS;][]{MScavallini06, MSreardon08} instrument at the National Solar Observatory Dunn Solar Telescope. The observations consist of a long time series (more than three hours) of full-Stokes high-spatial and temporal resolution spectropolarimetric scans (21 spectral points) of the Fe I 617.3 nm and Ca II 854.2 nm spectral lines of AR12546, one of the largest sunspots emergent onto the solar surface over the last $20$ years. The spectral sampling is $20$ m$\angs$ and $60$ m$\angs$ for the Fe I 617.3 nm and Ca II 854.2 nm spectral lines, respectively. The cadence of the reduced data is 48~s, and the data set was acquired during stable seeing conditions for $184$ min starting at $13:39$ UTC. At the beginning of the observation AR12546 was very close to disk center [$ 7^\circ$ S, $2^\circ$ W]. The adaptive optics system \citep[AO,][]{2004SPIE.5490...34R} was locked and running on the center of the sunspot and the integration time was set to $80$ ms. The theoretical diffraction-limited spatial resolution of the data is governed by the telescope aperture and the observed wavelengths, yielding $0.16$~arcsec and $0.23$~arcsec for the Fe $617.3$ nm and Ca $854.2$ nm spectral lines, respectively. During the acquisition period, the solar active region of interest (AR12546) was located at disk center.\\
It is worth mentioning that the long duration of the data set, together with the fast acquisition cadence, is ideal for the study of oscillatory phenomena with periods ranging from $2$ min (i.e. Nyquist sampling limit), to $\sim 2$ hours. Moreover, the availability of simultaneous observations at the photosphere and chromosphere allows us to search for signatures of propagation of magnetic field disturbances in circular polarization ($CP$) measurements, including an examination of the locations where this takes place. \\
In addition to the standard calibration procedures (i.e. flat fielding, dark subtraction, and polarimetric calibrations), the images obtained where restored with MOMFBD \citep{MSnoort05} techniques in order to limit the effects of the residual atmospheric aberrations left over by the AO system. \\
In order to provide context over a larger FoV, in figure \ref{Fig:context} we show the continuum intensity, the field strength, and inclination maps derived from the HINODE SOT/SP near-simultaneous observations of the same region in the Fe I $630.1$~nm and $630.2$~nm spectral lines \citep{springerlink:10.1007/s11207-008-9174-z}. The maps are part of the SOT/SP level 2 data products \citep{2013SoPh..283..601L}, which are outputs from spectral line inversions using the MERLIN code \citep{data}, and were downloaded from the HAO-CSAC (Community Spectro-polarimetric Analysis Center) data center\footnote{https://www2.hao.ucar.edu/csac}. Here, we note that this sunspot is very peculiar for both its size and magnetic field strength. Indeed, the field strength at the center of the umbra reaches values exceeding $4000$ G. This is an uncommon magnetic field value for a sunspot \citep[see, for example,][for a detailed analysis of the distribution of magnetic field strengths of sunspots]{2015A&A...578A..43R}. However, it is worth mentioning here that very recently \citet{2041-8205-852-1-L16} have reported field strengths in excess of 6000 Gauss.  In the same maps, the dot-dashed box represents the IBIS FoV. Due to the large dimensions of the magnetic structure, the IBIS FoV nearly approximates the umbra diameter in the $x$-direction, while it includes part of the penumbra in the $y$-direction. In the same figure, we show an IBIS intensity image obtained in the core of the Fe 617.3 nm spectral line, the simultaneous chromospheric counterpart acquired in the core of the Ca II $854.2$ nm spectral line, whose average height of formation is in the range $800-1000$ km \citep{2006ASPC..354..313U}, and the corresponding $CP$ maps deduced from these IBIS observations. Also in this case, we note that the signal at the center of the umbra is very weak ($\sim 100-200$ DNs). As previously described, the integration time of the IBIS instrument was set at $80$~ms. This value was chosen so as to almost completely freeze the effects of the residual seeing, thus allowing the application of deconvolution techniques. For this reason, unfortunately, the integration time could not be extended to allow the increase of the photon flux at the center of the umbra. \\
In this study we measured the circular polarization signal on a pixel-by-pixel basis by considering the amplitude of the Stokes-V profile:
\begin{equation}
CP = \frac{|V_{max}|}{I_{cont}} \cdot sign(V_{max}),
\end{equation}
where $V_{max}$ is the maximum amplitude of the Stokes-V spectral profile, and $I_{cont}$ the local continuum intensity. The choice of this particular definition of $CP$, instead of the more common integral of the Stokes-V profile, suits our goal to perform phase lag analyses between the photospheric and chromospheric channels. Indeed, the adopted definition results in the best spatial resolution when compared to other methods, since consecutive images during each scan of the corresponding spectral line are not summed up, preventing the smearing of small-scale details. Besides, the above definition also ensures the maximization of the spectral coherence between different layers.\\
The spectropolarimetric sensitivity of IBIS is very high and was estimated in \citet{2010ApJ...723..787V} as $10^{-3}$ the continuum intensity level. \\
At the center of the studied sunspot, we observe a saturation of the polarization signals in the photosphere. This can be explained as a combination of different effects; namely the low signal-to-noise ratio caused by the intrinsic photon flux at the center of the umbra, challenging the dynamic range of the detector \citep{2014SoPh..289.3531C}, and the saturation of the magnetic field sensitivity \citep[see for instance][]{2013A&ARv..21...66S}, which occurs when Zeeman splitting becomes comparable with the line width and is typical of spectral lines with large Landé factors such as Fe I $617.3$ nm ($g=2.5$). For these reasons, the pixels at the center of the umbra are masked out in the maps. However, we note that this region is not included in any of the analyses of this work.\\
We studied the $CP$ perturbations by applying an FFT (Fast Fourier transform) coherence and phase analysis to $CP$ signals at the two atmospheric heights sampled by the observations. In addition, we also undertook phase correlation analyses between $CP$ and intensity oscillations to attempt to identify a possible coupling between them.\\
In order to ensure the reliability of our results and their statistical significance, all the phase-lag diagrams reported in this work only include phase measurements that demonstrated coherence levels larger than $95\%$. Indeed, any phase estimate between signals with a small coherence should not be considered physically meaningful. We note that a coherence threshold of $95\%$ corresponds to a very large confidence level, ensuring accurate relationships (e.g. coupling) between the two different diagnostics are investigated.

  \begin{figure*}
  \centering
  (a)\includegraphics[trim=1cm 0cm 0cm 0cm, clip, height=4.5cm]{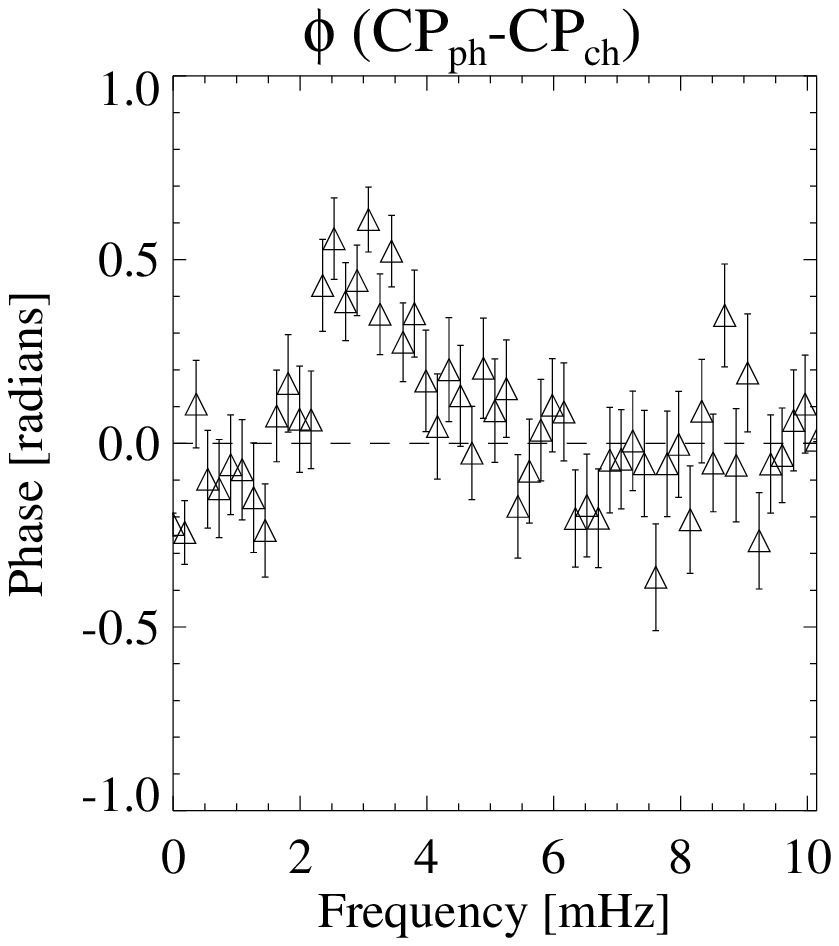}
  (b)\includegraphics[trim=1cm 0cm 0cm 0cm, clip, height=4.5cm]{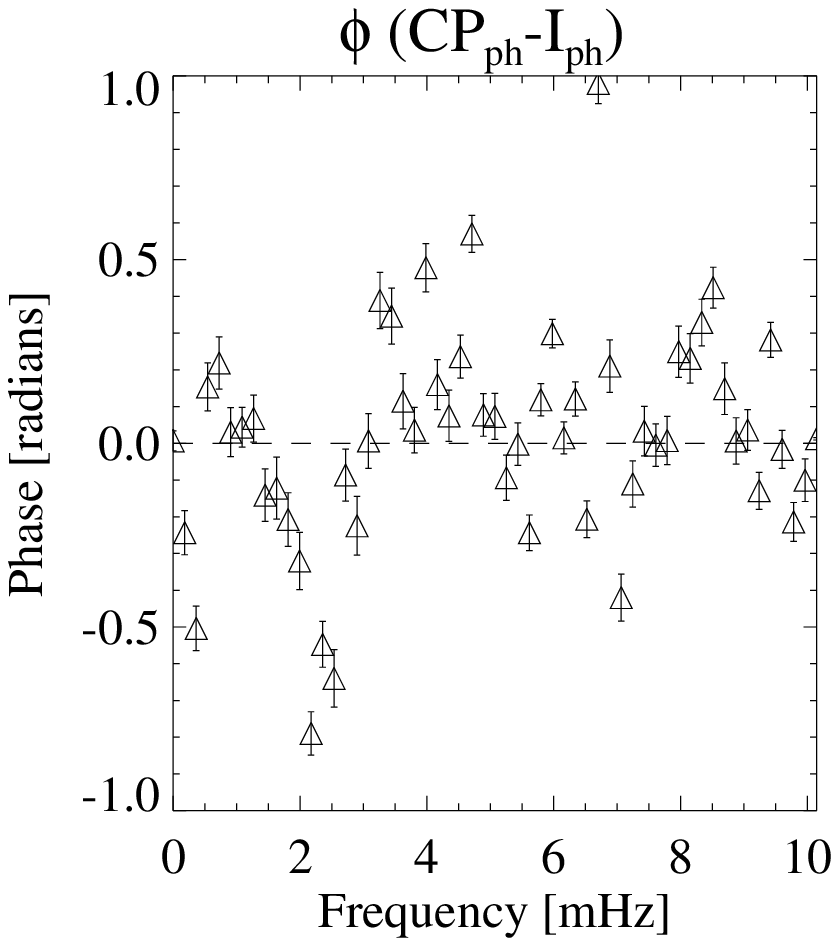}
  (c)\includegraphics[trim=1cm 0cm 0cm 0cm, clip, height=4.5cm]{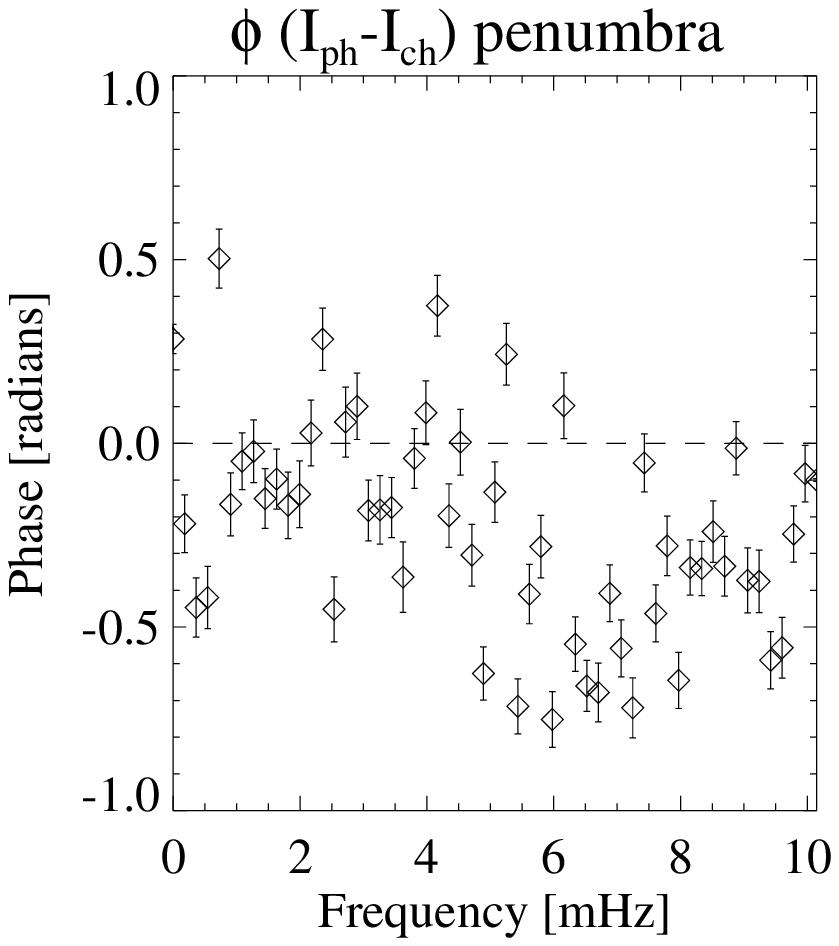}
  (d)\includegraphics[trim=1cm 0cm 0cm 0cm, clip, height=4.5cm]{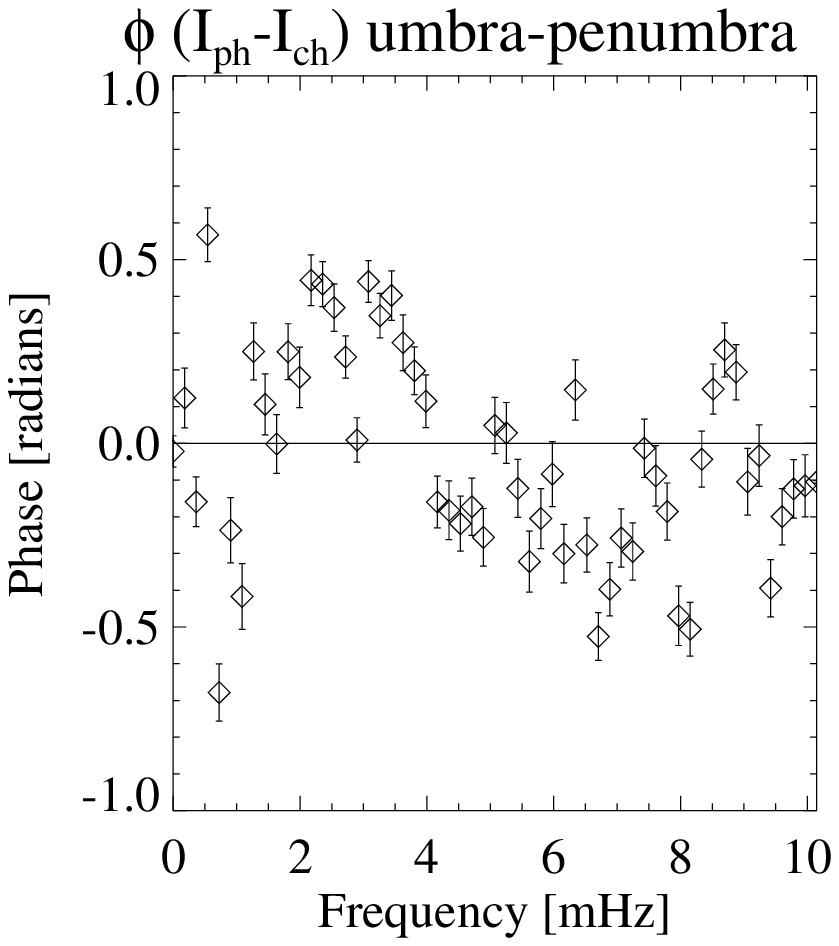}
  \caption{(a) Phase lag diagram between the $CP$ signals in the photosphere and chromosphere computed in the annular region highlighted by the dashed lines in panel (b) of Fig. \ref{Fig:phasemaps}. (b) $CP$-Intensity phase diagram in the annular region of Fig. \ref{Fig:phasemaps} panel (b) estimated in the photosphere. All plots are obtained by considering only those phase values for which the coherence is larger than $95\%$ in each frequency bin. (c) Phase lag diagram of intensity perturbations in the core of the Fe I $617.3$ nm and Ca II $854.2$ nm spectral lines in the penumbra outside the annular region of panel (b) Fig. \ref{Fig:phasemaps}. (d) Phase lag diagram between intensity perturbations in the core of the Fe I $617.3$ nm and Ca II $854.2$ nm spectral lines. The diagram is obtained in the annular region highlighted by the dashed lines in panel (b) of Fig. \ref{Fig:phasemaps}.}
  \label{Fig:phase_diagrams}
  \end{figure*}

It is worth noting here that the time series of certain diagnostics in the solar atmosphere may be the result of the superposition of different oscillations and processes. There might be different modes present at the same time, and those showing a coupling (i.e. a particular phase lag) between, for example, different heights may have a much smaller amplitude compared to the dominant peak of the phase spectrum.
In this regard, the coherence analysis can be seen as a filtering technique, which highlights the mere coupled components of two signals, even if these components have a small amplitude compared to other dominant peaks of the power spectrum. 

\section{Results}
\subsection{CP-CP phase}
In Fig. \ref{Fig:phasemaps}, we show the $CP$ phase lag map (panel a) and its related coherence map (panel b) in the $3$ mHz band ($0.7$ mHz bandwidth) as obtained from our FFT coherence analysis of the $CP$ signals at the two atmospheric heights sampled by the Fe I $617.3$ nm and Ca II $854.2$ nm spectral lines. The map is obtained by averaging the phase values in this range. Our reasoning in choosing this frequency band will become apparent later in this section. Here, we note that the phase lag map displays a circular area of positive values of the phase at the umbra-penumbra boundaries. In our sign convention a positive phase means that the photospheric signal is lagging behind the chromospheric counterpart (i.e. downward propagation of the perturbation). At the same location where a positive phase is observed (panel a), we also observe a large value of coherence (see panel b). 
This fact ensures the reliability of the phase estimates themselves and can be regarded as strong evidence of coupling between the two $CP$ signals at the two heights in the solar atmosphere. We remark that the phase lag was corrected for the time lag introduced by the instrumental sequential scanning of the spectral lines (i.e. the two spectral lines are scanned sequentially within $48$ s). It is worth noting that any instrumental effect would not result in an annular spatial distribution of the propagating disturbances as shown in the panel (a) of Fig. \ref{Fig:phasemaps}.\\
The phase and coherence maps shown in this figure are the result of an average over four frequency bins or, equivalently, $3.0\pm0.7$~mHz. This means that the coherence can be lower than the maximum value found in a single frequency bin, due to the intrinsic dispersion of the phase measurements themselves.  Despite this, the average coherence is quite large in the annular region, with frequency-averaged values exceeding $0.7$. Indeed, as we will see, the maximum coherence at each frequency can be larger than this value and, in some cases, exceeds $0.95$.\\ 
In order to investigate further the phase spectrum between the photospheric and chromospheric signals, in panel (a) of Fig. \ref{Fig:phase_diagrams}, we plot the phase spectrum obtained by considering only those pixels where the coherence is above $95\%$ in the annular region highlighted by the dashed lines of Fig. \ref{Fig:phasemaps} (panel b) and in each frequency bin of the spectrum. It is worth noting that, in contrast to the phase maps of Fig. \ref{Fig:phasemaps} obtained by averaging the phase over four spectral bins, the phase diagrams here are obtained by over-plotting on the same graph only those phase measurements with a coherence larger than $0.95$. For this reason, as was discussed previously,  lower coherence values are found in the phase maps, albeit large enough ($0.7-0.8$) to be considered a robust confidence level. The $CP$ phase spectrum shows a distinct positive peak at $2.5-3$ mHz, which is thus the frequency range considered to create the average maps shown in Fig. \ref{Fig:phasemaps}.

\subsection{CP-I phase}
In order to investigate if the observed $CP$ disturbances are associated with intensity oscillations, we also studied the relationship between the photospheric $CP$ oscillations and the intensity fluctuations in the core of the Fe I $617.3$ nm spectral line.
The resulting phase and coherence maps at the same frequency band are shown in Fig. \ref{Fig:phasemaps}, panel (c) and (d), respectively. We do not observe an annular region of positive phase values as clear as in the case of the $CP$-$CP$ phase map of panel (a). Of course, here the phase relationship is determined at a single atmospheric height, with non-zero phase lags expected between particular MHD wave modes in $CP$-$I$ measurements \citep{2009ApJ...702.1443F, 2015SSRv..190..103J}. However, the coherence map (panel d) does show large values at the umbra-penumbra boundaries and in the penumbra. This latter observation means that there exists a coupling between the $CP$ and intensity perturbations (i.e. there is an intensity perturbation corresponding to a co-spatial $CP$ perturbation).\\
The above coupling can be better seen in the phase diagram shown in panel (b) of Fig. \ref{Fig:phase_diagrams}, derived from  considering only those phase measurements with a coherence larger than $95\%$. This diagram is obtained in the annular region, including the umbra-penumbra boundaries, and highlighted by the dashed lines in Fig. \ref{Fig:phasemaps} (panel b). In this diagram we see that, although $CP$ perturbations are accompanied by intensity perturbations, these are not necessarily in phase. Indeed, at the same frequency and spatial location where we observe propagating circular $CP$ disturbances, we find negative phase values between $CP$-$I$ fluctuations at photospheric heights. A negative phase in our sign convention means that intensity is lagging behind $CP$.\\



\subsection{I-I phase}
In the previous section we have found that, although there exists a coupling between $CP$ and intensity, they are not in phase, and that the intensity perturbations are delayed with respect to the $CP$ fluctuations. In order to complete this picture, we have studied the vertical propagation of intensity disturbances in the core of the two spectral lines sampled by IBIS.  This was done in the penumbra (i.e. the region outside the dashed lines of Fig. \ref{Fig:phasemaps} panel (b)), and at the umbra-penumbra boundaries (i.e. the annular region shown in Fig. \ref{Fig:phasemaps} panel (b)). In particular, the penumbra was chosen as a reference and for comparison of the propagating regime observed in the umbra-penumbra boundaries. The results of this analysis are shown in panels (c) and (d) of Fig. \ref{Fig:phase_diagrams}, where we plot the $I$-$I$ phase diagram in the penumbra and at the umbra-penumbra boundary.\\
In the penumbra, we observe the upward propagation (negative phase) of the intensity oscillations for frequencies larger than $3-3.5$ mHz. This is consistent with the presence of a cut-off frequency and upwardly propagating magneto-acoustic waves, which is consistent with the findings for running penumbral waves \citep{2013ApJ...779..168J}. However, in addition to this upward propagating regime for frequencies larger than $3-3.5$ mHz, at the umbra-penumbra boundary we also observe a second positive-phase component at $2.5-3$ mHz that was not detected in the penumbra (panel c of the same figure).  Even in this case, the phase diagrams were obtained by only considering the phase measurements corresponding to a coherence larger than $95\%$.

  \begin{figure}
  \centering
  \includegraphics[trim=0.cm 0cm 0cm 0.cm, clip, width=5cm]{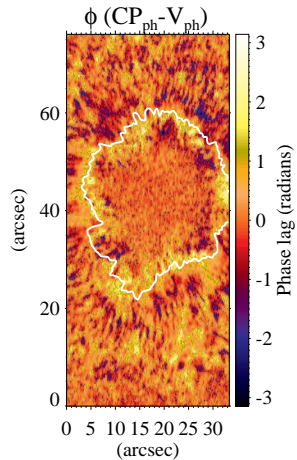}
  \caption{Phase lag map between $CP$ and LoS velocity measured in the photospheric Fe I 617.3 nm data.}
  \label{Fig:CP_LOS}
  \end{figure}

\section{Summary and discussions}
In this work, by taking advantage of a long duration spectropolarimetric data set (more than three hours) sampling two different heights in the solar atmosphere,  we have studied the propagation of $CP$ disturbances in a large and symmetric sunspot, alongside their coupling with other physical quantities. The main findings of our work can be summarized by the following:
\begin{itemize}
 \item downward propagating $CP$ disturbances, at $2.5-3$ mHz, are detected with a high confidence level at the umbra-penumbra boundary (see panel (a) and (b) of Fig. \ref{Fig:phasemaps});
\item at the same location where propagating $CP$ fluctuations are identified, we also detected intensity oscillations that lag behind their corresponding $CP$ perturbations (see panel (c) of Fig. \ref{Fig:phase_diagrams});
\end{itemize}
 
Several authors have reported oscillations of the Stokes profiles at umbra-penumbra boundaries in sunspots, attributing them to temporal variations of the observed magnetic field \citep[e.g., ][]{1998A&A...335L..97R, 1999SoPh..187..389B, 2000A&A...355..347Z, 2018arXiv180300018H}. Based on phase lag analyses between different observables restricted to a single height in the solar atmosphere, the same authors argued that these oscillations could be associated with real magnetic oscillations, excluding instrumental or opacity effects. However, this is a controversial opinion. Indeed, \citet{2002A&A...392.1095S} have underlined that several observational artifacts can introduce oscillations in the measurements of the magnetic field vector, e.g. seeing effects, instrumental cross-talk from velocity, and opacity effects.\\
In our study, artifacts from instrumental cross-talk from velocity can be ruled out by the phase delay measured between the magnetic perturbations and the velocity itself that is displayed in Fig. \ref{Fig:CP_LOS}. Besides, we note that the frequency, spatial coherence and vertical propagation of the disturbances are all aspects that make seeing effects unlikely to contribute to the obtained results. However, as already stressed in \citet{2000ApJ...534..989B}, a definitive interpretation of the oscillations in the magnetic field would only be possible through the study of the fluctuations of atmospheric diagnostics and their stratification in terms of geometrical heights. Unfortunately, this is not a trivial task, as it requires the identification of a reference height, which is also subject to opacity effects. However, by comparing a theoretical model to observations, \citet{2003ApJ...588..606K} have argued that the observed fluctuations of the magnetic field are the result of a mixture of intrinsic magnetic oscillations, though rather small (a few G), and time-varying opacity effects due to magneto-acoustic waves. In particular, they found that towards the edges of the umbra it is impossible to reproduce the observations without including the intrinsic oscillations of the magnetic field that characterize the fast MHD mode, something that has recently been further quantified by \citet{Grant:2018vfa}. Once again we stress that, in contrast to the power spectrum where small coherent signal might be hidden in other dominant components, the phase lag and coherence analysis acts as a robust "\textit{de-noising}" technique, highlighting small-amplitude, yet coherent (between different layers) signals. 
In this regard, our techniques may identify only the coherent part of downwardly propagating $CP$ perturbations, regardless of their amplitudes.\\
Our analysis also reveals coupled intensity perturbations lagging behind their corresponding $CP$ fluctuations. This is in contrast with \citet{1999SoPh..187..389B}, who found no statistically significant relationship between intensity and magnetic field variations. The coherence between the two signals in their case was in fact below $0.5$, therefore, it was too small to draw conclusions on a possible relationship between the two quantities. However, we note that a lower coherence value may have resulted, in their case, by the lower quality of the dataset employed by the authors with respect to the one studied here. Indeed, no AO was available, and this resulted in larger dynamic optical aberrations that may have reduced the coherence between the different quantities investigated. In this sense, our results regarding the coupling of $CP$ and intensity oscillations are not in contrast to those obtained by \citet{1999SoPh..187..389B}, but simply underline the necessity of very good seeing conditions for accurate phase lag analyses.\\
In the case of opacity effects, where the magnetic field fluctuations are a consequence of the vertical magnetic field gradient (i.e. $dB/dz$), we expect the intensity oscillations to be out-of-phase with the $CP$ fluctuations. This is not the case for our observations, thus the presence of genuine magnetic fluctuations appears to be a reasonable conclusion. \\
In this regard it is worth recalling that, very recently, \citet{2018arXiv180301737J} have deeply investigated magnetic oscillations in the umbra and penumbra of a sunspot through non-LTE spectropolarimetric inversions at chromospheric heights, and found magnetic field variations that are not in agreement with opacity effects.\\ 
We note that the phase diagram of intensity oscillations between two layers shows the coexistence of two modes at the umbra-penumbra boundary; upward propagating waves for frequencies above the cut-off, and downward propagating oscillations at $2.5-3$ mHz. The latter are also accompanied by downward propagating $CP$ oscillations. \\
In an attempt to explain the presence of magnetic oscillations at the umbra-penumbra boundary, \citet{1999ESASP.448..417Z} have shown theoretically that these oscillations can be ascribed to either slow surface or body modes, although the first option was soon ruled out since it is expected that a surface mode would appear in a thin layer with an estimated width of $\sim 100$ km. In our case, the width of the region showing the presence of propagating $CP$ oscillations is of the order of $10$ arcsec, or equivalently $7500$ km.  Nevertheless, the absence of propagating disturbances in the umbra of the sunspot seems to agree with the surface wave scenario. Here, we note that the detection of downward propagating surface disturbances in the sunspot is independent of the possible contamination of opacity effects on the $CP$ signals. Indeed, as mentioned previously, it was demonstrated that these opacity effects would be the signature of magneto-acoustic perturbations anyway. Recent work by \citet{2018ApJ...857...28K} provided a framework for directly detecting surface and body modes in pores, which are simpler structures in comparison to sunspots. The inclination of penumbral fields means that these techniques are more complicated to apply to our data to verify our belief that these oscillations are surface modes. Although it is outside the scope of this current study, in future work we will look to adapt these techniques to work with sunspots and to determine conclusively if these oscillations are surface modes.\\ 

\acknowledgments
This work has been partly supported by the "Progetti di ricerca INAF di Rilevante Interesse Nazionale" (PRIN-INAF) 2014 and PRIN-MIUR 2012 (prot. 2012P2HRCR) entitled "Il sole attivo ed i suoi effetti sul clima dello spazio e della terra" grants, funded by the Italian National Institute for Astrophysics (INAF) and Ministry of Education, Universities and Research (MIUR), respectively. The National Solar Observatory is operated by the Association of Universities for Research in Astronomy under a cooperative agreement with the National Science Foundation. IBIS has been built by the INAF-Arcetri Astrophysical Observatory, with the support of the Dept. of Physics and Astronomy of the University of Florence, and the Dept. of Physics of the University of Rome - Tor Vergata. It is operated and supported by INAF in collaboration with the US National Solar Observatory. RE thanks the Science and Technology Facilities Council (STFC, grant numbers ST/M000826/1, ST/L006316/1) for the support to conduct this research. SJ acknowledges support from the European Research Council (ERC) under the European Union’s Horizon 2020 research and innovation program (grant agreement No. 682462) and from the Research Council of Norway through its Centres of Excellence scheme, project number 262622. DBJ would like to thank the STFC for the award of an Ernest Rutherford Fellowship, in addition to a dedicated standard grant which allowed this project to be undertaken. DBJ also wishes to thank Invest NI and Randox Laboratories Ltd. for the award of a Research \& Development Grant (059RDEN-1). PHK would like to thank the Leverhulme Trust for the award of an early career fellowship. Hinode is a Japanese mission developed and launched by ISAS/JAXA, collaborating with NAOJ as a domestic partner, NASA and UKSA as international partners. Scientific operation of the Hinode mission is conducted by the Hinode science team organized at ISAS/JAXA. This team mainly consists of scientists from institutes in the partner countries. Support for the post-launch operation is provided by JAXA and NAOJ (Japan), UKSA (U.K.), NASA, ESA, and NSC (Norway).

\bibliographystyle{yahapj}

\end{document}